# From Social Network to Semantic Social Network in Recommender System

Khaled Sellami[1], Mohamed Ahmed-Nacer[2] and Pierre Tiako[3]

[1] Computer Science Department, A/Mira University of Bejaia
Route de Targua Ouzemour, 06000, Bejaia, Algeria
*Khaled.sellami@univ-bejaia.dz*

[2] LSI Laboratory, USTHB University
BP32 El Alia, Bab Ezzouar, Algiers

[3] Center for IT Research, Langston University
Oklahoma, 73050, USA

**Abstract**
Due the success of emerging Web 2.0, and different social network Web site such Amazon, and movie lens, recommender systems are creating unprecedented opportunities for to help people browsing the web when looking for relevant information and making choices. Generally, these recommender systems are classified in three categories: content based, collaborative filtering, and hybrid based recommendation systems. Usually, these systems employ standard recommendation methods such as artificial neural networks, nearest neighbor, or Bayesian networks. However, these approaches are limited compared to methods based on web applications, such as social networks or semantic web. In this paper, we propose a novel approach for recommendation systems called semantic social recommendation systems that enhances the analysis of social networks exploiting the power of semantic social network analysis. Experiments on real-world data from Amazon examine the quality of our recommendation method as well as the performance of our recommendation algorithms.

## 1. Introduction

The prevalent use of computers and Internet has enhanced the quality of life for many people, tasks that were once done mostly through physical/human interactions, such as banking, shopping, or communication, can now be done online a seemingly simpler and better alternative. Also, with rapidly growing amount of information in the web, it is difficult to find needed information quickly and efficiently. That is where the recommender systems come in as a special type of information filtering. Nowadays many applications have used recommender systems; especially in the e-commerce domains such as http://www.amazon.com (see an example in Figure 1) where a failure recommendation could cause great losses of time, effort, and money. Our objective is to review a solution to surpass the defects of failure recommendation, by presenting semantic-social recommendation approaches. The idea here is to combine two important aspects, the social aspect by using social network analysis measures and the semantic aspect by using the semantic similarity measures.

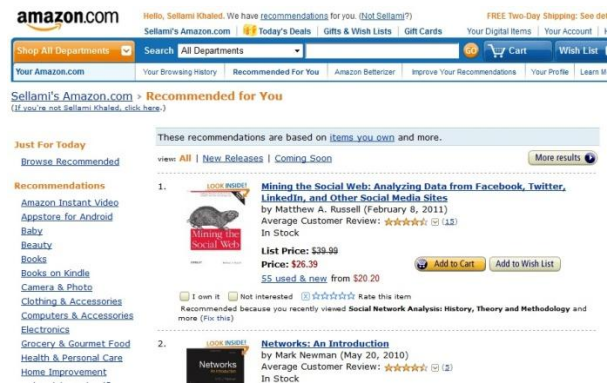

**Fig 1: Amazon recommends products to customers by customizing CF systems**

Recommender systems has three main categories [2]: content-based [5] where the users are recommended with items that are similar to those that they liked in the past, collaborative-filtering or social recommendation [19] where the recommendation depends on the user's neighbors' opinions and not on the item itself, and hybrid recommendation that combines the content-based and social based recommendation methods [11].

The paper is organized as follows: section 2 presents related work, section 3 details the new recommendation model proposed, section 4 explains our obtained results, and section 5 concludes and discusses future work.

## 2. Background knowledge and related works

The approach described in this paper relies on a combination of social network analysis and semantic web for semantic social recommendation. In this section, we explore related works in recommendation systems using these techniques. We also highlight the originality of the approach we propose with respect to the state of the art.

### 2.1 Classification criteria for recommender systems

The recommendation problem is defined as follows [2]: let $C$ be the set of all users and let $S$ be the set of all possible items that can be recommended. Let $u$ be an utility function that measures the usefulness of item $s$ to user $c$. $u : C \times S \rightarrow R$, where $R$ is a totally ordered set (non-negative integer or real numbers within a certain range). Then for each user $c \in C$, we want to choose such item $s' \in S$ that maximizes the user's utility. $\forall\ c\ \in\ C$, $s'_c = argmax_{s \in S} u(c,s)$.

In the recommendation systems, the utility $u$ refers to the rating. Each elements $c$ of the user space $C$ could be defined with a profile that contains the users' characteristics (id, name, age . . .). Each element $s$ of the items space $S$ is also defined with a set of characteristics. Traditionally, filtering and recommender systems were classified into three main categories relative to the filtering technique used [2]: content-based recommender systems [5], collaborative-filtering or social based recommendation [19], and hybrid recommendation systems [11].

In content-based recommender systems, users are recommended with items that are similar to those that they liked in the past [5]. The utility $u(c, s)$ of item $s$ for user $c$ is estimated based on the utilities $u(c, s_i)$ assigned by user $c$ to items $s_i \in S$ where $s_i$ are similar to item $s$ [2]. Generally, content based recommender systems depend on three main processes: content analyzer, profile learner and filtering components [26]. The content analyzer is used to extract information (keywords, concepts, etc) that represent items, and to extract users reactions towards these items. The profile learner is used to learn users' preferences, from their past reactions towards items, in order to construct and update user profile. Filtering components matches user profile with items characteristics to accomplish the recommendation.

In collaborative filtering recommender systems, recommendation is based on the user's neighbors' opinions not on the item itself [28]. The utility $u(c, s)$ of item $s$ for user $c$ is estimated based on the utilities $u(c_j, s)$ assigned to item $s$ by those users $u(c_j, s)$ who are similar to user $c$ [19].

Collaborative filtering recommender systems have three types: item-based, user-based and item-user-based [31]. In user-based collaborative filtering, a user $C$ who is interested in item $x$ will be interested in item $y$, if there are other users. These users are similar to the user $C$ and they are interested in the items $x$ and $y$ [28]. In item-based collaborative recommender system, if a user $C$ likes item $x$, and the item $x$ is similar to the item $y$ according to the opinion to other users. Then item y should be recommended to the user C[29].

Hybrid recommender systems combine the characteristics of content-based and collaborative filtering methods for avoiding some limitations and problems of pure recommender systems, like the cold-start problem. The combination of approaches can proceed in the following different ways [2]:
1) Separate implementation of algorithms and joining the results.
2) Utilize some rules of content-based filtering in collaborative approach.
3) Utilize some rules of collaborative filtering in content based approach.
4) Create a unified recommender system that brings together both approaches.

However, another classification criterion of RS may be considered. For example, Depending on the information filtering method, there are (1) passive filtering systems [27] when a single recommender is generated for all system users, and (2) active filtering systems [8] where the recommendation is generated from the user's recommendation history to generate new customized recommendations.

There are also distinctions to be made between centralized systems (when the product descriptions and user profiles are stored in a centralized Server) and non-centralized Systems (generally developed on P2P networks).

We can also classify RS by considering the way user preferences are obtained and distinguish between explicit data collection systems (when the user is asked to voluntarily provide their valuations) and implicit data collection systems (where the system user is monitored).

The list below gives us an idea of the range of kind of classification criteria we can find in the literature. This list is not exhaustive, however, we are interested here on analyzing the state of this subject for a specific category of recommendation system that consist of the most future line of research in recent years: semantic social recommender systems

### 2.2 Recommendation systems

The main idea of collaborative filtering recommender systems is to capture the user's tastes, compute the similarity between users, and predict the recommendations. Generally all the collaborative filtering algorithms have the main principals, but they differ in the way of computing the similarity between users.

Early generation collaborative filtering systems, such as GroupLens [28], propose Newsnet; the article recommender system. Which is a user-based, and uses Pearson r correlation coefficient to compute the similarity or weight between users and make predictions or recommendations according to those calculated similarity values. Later, Grouplens implemented this algorithm on Usenet news [17].

In [30] authors introduced a personalized recommender system called Ringo, which recommends music and artists to users. For this system the authors implemented and compared four CF algorithms. These algorithms are: the mean squared differences algorithm; which measures dissimilarity between users, the Pearson r algorithm, the constrained Pearson r algorithm and the item-based CF algorithm. Their results showed that the constrained Pearson algorithm gives the best results.

In [18] Spearman ranking correlation coefficient as another recommendation measure is proposed. Spearman correlation is the same to Pearson correlation, but instead of handling the ratings the algorithm handles the ranking of the ratings. These results proved that Spearman ranking correlation performs as well as Pearson correlation.

In [3] authors proposed an intelligent recommendation algorithm called IRA. This algorithm is a graph based collaborative filtering recommendation algorithm, where users are connected via directed graph. The nodes of this graph represent users while the directed edges of this graph represent the horting and predictability relation between these users; horting and predictability relation is mathematically defined in [3]. The algorithm recommends the item j to the user I by computing the shortest path in its entirely between the user i and group of users. Each user in this group should have common rated items with the user i and should have already rated the item j. In this algorithm the author proposed the breadth first search algorithm to compute the shortest paths between users.

In [22] the authors proposed Movie recommender system. In this system three graphs have been defined, the first graph is the bipartite graph. Its nodes are divided into two sets the people set P and the movie set M and the edges E are created between P and M and represent the ratings and viewing preferences between P and M. The second graph is the collaboration network graph which is a one-mode projection graph between the users; two users will have collaboration connection between them, if they have at least one movie in common. The third graph is the recommender graph which is a sum of the social collaboration graph and the bipartite graph. In order to give the recommendation, shortest path algorithm is applied on the recommender graph.

The limitation of the aforementioned works is the tight coupling with the collaborative filtering recommendation. Even if there are several graph based recommender systems, these recommender systems never employ the social network analysis measures in the recommendation algorithm. For that, we propose to involve the social network analysis measures in the recommendation algorithm. Furthermore, we also propose to involve the user's semantic preferences in this recommendation algorithm, in order to have a semantic-social recommendation algorithm.

### 2.3 Social network

Social Networks are networks in which vertices represent users, and edges represent links (social relations such as friendship and co authorship) among these users [24]. Social network analysis is the study of social networks by understanding their social entities, the people and their relationships, examples considered indirectly as forms of social networks are: telecommunications, electronic mail, and electronic chat messengers (such as Skype, Google Talk or MSN Messenger). Actually, social network analysis measures are used to study the structural properties of the social network [24, 14]; the **density** indicates the cohesion of the network. The **centrality** highlights the most important actors of the network and three definitions have been proposed, the **degree centrality** is based on the average length of the paths (number of adjacent edges). The **closeness centrality** is based on the average length of the paths (number of edges) linking a node to others and reveals the capacity of a node to be reached. The **betweenness centrality** focuses on the capacity of a node to be an intermediary between any two other nodes.

Furthermore, due to the recent evolution of social networks, social recommender systems are becoming more common such as: (a) Finding the user's best co-workers in a social network [25]. (b) Recommending friends, using graph based algorithms such as random walk [21]. (c) Proposing music in a social network of connected artists [10]. (d) Tagging based recommender system for recommending photos [1]. Bookmarking uses a personalized tag recommendation system for users of bookmarking sites using text mining similarity measures [9]. [4] Presenting a Facebook group recommender system, by using hierarchical clustering and decision tree techniques. Also, Facebook application has been proposed in [6] to find colleagues who can work in similar projects.

### 2.4 Semantic Social network

As we have seen, the use of software instead of users in the information filtering has certain weaknesses: i) how to represent information complicates communication among agents and between agents and users, ii) reuse of information represented heterogeneously becomes too complicated.

With the arrival of the Semantic Web [7], these deficiencies are mitigated by the improvement and enrichment of the representation of information through the application of these technologies. Semantic Social Network is the composition of two types of technologies: semantic web technology [7] and social networks technology [14]. The first research question about the possibility of having a semantic social network was presented in 2002 [16]. Later in 2004 Stephen Downes [13], has proposed new type of Internet as a network within a network to reshape the Internet that we know. This type of Internet is based on merging semantic web technology and social network [14].

In [14], authors have proposed semantic social network analysis model semSNA, where social data are presented in RDF [1]. Then social network analysis features e.g. closeness centrality, betweenness centrality and graph annotations are computed using SPARQL[1]. In [20], authors have used the social network analysis (SNA) for analyzing ontology and semantic web, they have applied some of social network analysis techniques on two different ontology's SUMO [2] ontology, and SWRC [3] ontology. In recent years many search has focused on the analysis of the semantic social networks and that propose various solutions in different fields, basically, they can be classified by way of representing the semantic aspect as: Semantic user profile in the social network, and Social Networking Ontologies.

#### 2.4.1 Semantic user profile in the social network

Semantic user profiles have become a key part of adequate social network. In [23], authors have presented a semantic social network, applied to the PUII (Program for the University Industry Interface). Its objective was to identify the employees' skills in a company and to deal with knowledge in online communities. In this project the semantic social network is based on: firstly, Meta data representation of users and resources. Secondly, information tailoring of user profile, using social network and ontology's, and finally, the semantic interoperability (Profile).

In [12], authors have used a multi-layered model to present the semantic social network, ontology has been presented as a semantic network of interrelated domain concepts, while user profiles have been described as weighted list of those concepts. User profiles have been clustered due to user's interests, and the similarity has been considered as a similarity measure between users and clusters,

#### 2.4.2 Social Networking Ontologies

The two most important achievements in build ontologies to classify social networking activities so far: FOAF[4] and SIOC[5].

---

[1] Semantic Web, W3C, http://www.w3.org/2001/sw/
[2] http://www.ontologyportal.org/
[3] http://ontobroker.semanticweb.org/ontologies/swrc-onto-2001-12-11.oxml
[4] http://www.foaf-project.org/

### FOAF

The Friend of a Friend (FOAF[4]) project, one of the largest projects in the semantic web, is a descriptive vocabulary built based on RDF and OWL, for creating a Web of machine-readable pages for describing people, the links between them and the things they create and do. It is accepted as standard vocabulary for representing social networks, and many large social networking websites use it to produce Semantic Web profiles for their users [15].

FOAF has the potential to become an important tool in managing communities, and can be very useful to provide assistance to new entrants in a community, to find people with similar interests or to gather in a single place, people's information from several different resources, decentralizing the use of a single social network service for example [15].

### SIOC

The SIOC[5] project (Semantically-Interlinked Online Communities), is an ontology for representing rich metadata from the Social Web in RDF/OWL, accepted by W3C. It aims to enable the integration of online community information (wikis, message boards, weblogs, etc).

### 3. Our Method

The recent emergence of semantic social networks (SSNs) gives us an opportunity to investigate the role of semantic social influence in recommender systems. The performance of semantic social recommender systems are based in one hand on knowledge base usually defined as a concept diagram (like taxonomy) or ontology and in another hand on social network analysis measures (like degree centrality, betweenness centrality, influence).

### 3.1 Our Hybrid Item-Based Similarity Matching Method

In this work, we have extended existing methods to develop a hybrid item-based similarity matching method.

**Item-Based Collaborative Filtering Multi-Attribute Similarity**:

We developed a multi-attribute rating scheme that allows users to rate an item along five attributes. The algorithm is described below:

*Step 1- Specify user preferences.* The user assigns the weight values ($W_A$) to each attribute along which similarities between information items are to be computed.

*Step 2- Compute the similarity between items with respect to every attribute (subject, performance, overall likability).*

For every attribute $A$, the similarity between information items $I$ and $J$ as given by [29]:

$$Sim_A(I,J) = \frac{\sum_{U \in Users}(R_A(U,I) - \bar{R}_A(U))(R_A(U,J) - \bar{R}_A(U))}{\sqrt{\sum_{U \in Users}(R_A(U,I) - \bar{R}_A(U))^2} \sqrt{\sum_{U \in Users}(R_A(U,J) - \bar{R}_A(U))^2}} \quad .(1)$$

---

[5] http://www.sioc-project.org/

Where $R_A(U,I)$ denotes the rating of user $U$ on item $I$ with respect to attribute $A$; $\check{R}_A(U)$ is the average rating of user $U$ as per attribute $A$.

Step 3- Compute the CF multi-attribute similarity between items

$$MultSim(I,J) = \frac{\sum_{A=1}^{3} W_A * Sim_A(I,J)}{\sum_{A=1}^{3} W_A} \quad (2)$$

**Item-Based Semantic Similarity:** In this method, we calculate the similarities between two items based on their semantic descriptions given in an ontology. The similarity between items $I$ and $J$ is based on the ratio of the common/shared RDF descriptions between $I$ and $J$ (*count_common_desc(I,J)*) to their total descriptions (*count_total_desc(I,J)*) as proposed by [32] and is given by:

$$SemSim(I,J) = \frac{count\_common\_desc(I,J)}{coun\_total\_desc(I,J)} \quad (3)$$

**Hybrid Item-Based Semantic-CF Similarity:** Using (Eq.**2** and Eq.**3**) we calculate the hybrid Semantic-CF similarity using a linear weighted approach as:

$$sim(I,J) = W_M * MultSim(I,J) + W_S * SemSim(I,J) \quad (4)$$

Where $W_M$ and $W_S$ are the weights assigned to CF multiattribute and semantic similarities respectively.

### 3.2 Semantic-Social Recommendation Algorithm

The recommendation algorithm is adopted and adapted from [32] shown in Table 1, we can define the algorithm input: as a product, and the algorithm output: as a group of customers (see table 1).

These customers are supposed to like the input product and to buy it.

**Table 1: steps of the recommendation algorithm**

| |
|---|
| 1. Start the search from the semantic social network.<br>2. Look for the customers with the highest influence node.<br>    (a) Compare the semantic profile of the customer with the semantic profile of the product (semantic comparison algorithm).<br>    (b) Compare the similarity degree between customer and product with a threshold.<br>    (c) If the recommendation degree is not enough<br>        i. start the search again from another customer.<br>        Go to (2).<br>    (d) Else if<br>        i. add the customer to the output group.<br>        ii. Move to the next customer.<br>        iii. Go to (a).<br>3. End |

In this section, we introduce the dataset that we use for this research, and present some interesting characteristics of this dataset. Our dataset is obtained from a real online social network Amazon.com . Our proposed algorithm depends on this dataset to build the semantic social network. For that, we have to build the semantic social network, then to apply the semantic social recommendation algorithm.

### 3.3 Performances of the semantic social network

To provide the recommendation algorithm, we built a semantic social network, from Amazon dataset, by two process: (a) Building the collaboration social network, each node of the social network represents a customer, and the edges represents the similarity between these customers (the similarity can be found, when two customers prefer same products with same ratings), (b) Building the semantic user profile, Amazon dataset has a conceptual presentation of products and users preferred products. A preliminary study on this collaboration social network yields the following results. The customers' number is 1974, the edges number is 125448, and the network density is 0,0664.

### 3.4 Performances of the recommendation algorithm

Give ten recommendation queries concerning a product that should be recommended to the most relevant customers in the semantic social network. The experimental results are listed in Table 3 and table 4. From these tables, we present for each query, the number of relevant customers, the computation time and the number of discovered nodes in table 2, the mean absolute error, the precision and the recall in the table 3.

**Table 2: For each item we apply recommendation query, this tables show the number of recommended customers, the recommendation time and the number of discovered nodes**

| Items | Customers number | Computation time | Graph coverage |
|---|---|---|---|
| Item1 | 2 | 3m31s | 361 |
| Item2 | 2 | 3m33s | 361 |
| Item3 | 23 | 3m11s | 361 |
| Item4 | 2 | 3m43s | 361 |
| Item5 | 15 | 3m33s | 361 |
| Item6 | 3 | 3m42s | 361 |
| Item7 | 2 | 3m16s | 361 |
| Item8 | 17 | 3m35s | 361 |
| Item9 | 4 | 3m45s | 406 |
| Item10 | 4 | 3m24s | 1123 |

**Table 3: For each item we apply recommendation query, this tables show the absolute error, the precision and the recall measures**

| Items | Precision | Recall |
|---|---|---|
| Item1 | 0,83 | 0,91 |
| Item2 | 0,53 | 0,91 |
| Item3 | 0,74 | 0,85 |
| Item4 | 0,61 | 0,74 |
| Item5 | 0,56 | 0,67 |
| Item6 | 0,55 | 0,56 |
| Item7 | 0,45 | 0,45 |
| Item8 | 0,55 | 0,69 |
| Item9 | 0,76 | 0,87 |
| Item10 | 0,65 | 0,69 |

### 2.4.3 Comparison Methods

As a comparison, we implemented the standard collaborative filtering algorithm as we described in Section 2. Then we notice that our recommendation algorithm provides a better precision / recall than the collaborative filtering algorithm, the computation time is better in the semantic-social recommendation algorithm (if the time of building or uploading the social network is not considered).

As a heuristic nature the proposed approach algorithm explores between 70% and 80% explored users.

## 3. CONCLUSION AND FUTURE WORK

Semantic social networks provide an important source of information regarding users and their relations enriched by knowledge base usually defined as an ontology. This is especially valuable to recommender systems. In this paper we proposed a semantic social recommendation algorithm which makes recommendations by considering a product recommendation to customers, which are connected via semantic social network, and we employs the social network analysis measures in the recommendation process, to benefit from the social relations between social network users. Our preliminary results by using Amazon dataset show a good computation time, good precision, and recall.

## 4. REFERENCES

[1] Adam Rae, Börkur Sigurbjörnsson, Roelof van Zwol. Improving tag recommendation using social networks Proceeding RIAO '10 Adaptivity, Personalization and Fusion of Heterogeneous Information. Paris France. 2010.

[2] G. Adomavicius and A. Tuzhilin. Toward the next generation of recommender systems: A survey of the state-of-the-art and possible extensions. Knowledge and Data Engineering, IEEE Transactions on, 17(6):734-749, 2005.

[3] C. C. Aggarwal, J. L. Wolf, K. Wu, and P. S. Yu. Horting Hatches an Egg: A New Graph-Theoretic Approach to Collaborative Filtering. In KDD '99: Proceedings of the fifth ACM SIGKDD international conference on Knowledge discovery and data mining, pp 201-212, San Diego, California, United States, 1999. ACM.

[4] E.-A. Baatarjav, S. Phithakkitnukoon, and R. Dantu. Group recommendation system for facebook. Pp 221-219. 2008.

[5] Balabanovic, M., and Shoham, Y., 1997. Fab: Content-based, collaborative recommendation. Communications of the ACM, 40:66-72, 1997

[6] S. D. Bedrick and D. F. Sittig. A scientific collaboration tool built on the facebook platform. AMIA ... Annual Symposium proceedings / AMIA Symposium. (2008):41-45, 2008.

[7] Berners-Lee, T., Hendler, J. y Lassila, O.. "The Semantic Web: A new form of Web content that is meaningful to computers will unleash a revolution of new possibilities". (2001) scientific American, May.

[8] Boutilier, C., Zemel, R.S. y Marlin, B. (2003). "Active collaborative filtering". Proc. Of the 19th Annual Conference on Uncertainty in Artificial Intelligence, pp. 98-106.

[9] A. Byde, H. Wan, and S. Cayzer. personalized tag recommendations via tagging and content-based similarity metrics. Proceedings of the International Conference on Weblogs and Social Media, (March 2007)

[10] P. Cano, O. Celma, M. Koppenberger, and M. J.Buldu. Topology of music recommendation networks. Chaos An Interdisciplinary Journal of Nonlinear Science, 16, 2006.

[11] M. Claypool, A. Gokhale, T. Miranda, P. Murnikov, D. Netes, and M. Sartin. Combining content-based and collaborative filters in an online newspaper, 1999.

[12] Cantador, I. y Castells, P. Multilayered Semantic Social Network Modeling by Ontology-Based User Profiles Clustering: Application to Collaborative Filterin. In Managing Knowledge in a World of Networks, pp. 334-34, 2006

[13] S. Downes. The semantic social network, February 2004, http://www.downes.ca/post/46 , published in The Learning Organization: An International Journal, Emerald Group Publishing Limited, Vol. 12 (2005) 411-417

[14] G. Erétéo, F. L. Gandon, O. Corby, and M. Buffa. Semantic social network analysis. CoRR, abs/0904.3701, 2009.

[15] J. Golbeck and M. Rothstein. Linking social networks on the web with foaf: A semantic web case study. In AAAI, pages 1138–1143, 2008.

[16] Henry M.Kim. Ontologies for the semantic web: Can social network analysis be used to develop them? Proceedings of the Conference on Computational Analysis of Social and Organizational Systems (CASOS), Pittsburgh, PA, June 21-3, 2002.

[17] J. A. Konstan, B. N. Miller, D. Maltz, J. L. Herlocker, L. R. Gordon, and J. Riedl. Grouplens: Applying collaborative filtering to Usenet news. Communications of the ACM, 40(3):77-87, 1997.

[18] J. L. Herlocker, J. A. Konstan, A. Borchers, and J. Riedl. An algorithmic framework for performing collaborative filtering. In SIGIR '99: Proceedings of the 22nd annual international ACM SIGIR conference on Research and development in information retrieval, pages 230-237, New York, NY, USA, 1999. ACM.

[19] T. Hofmann. Latent Class Models for Collaborative Filtering. In In Proceedings of the Sixteenth International Joint Conference on Artificial Intelligence, pages 688-693, 1999.

[20] B. Hoser, A. Hotho, R. Jäschke, C. Schmitz, and Stumme, Gerd. Semantic Network Analysis of Ontologies, 3rd European Semantic Web Conference, ESWC 2006, Budva, Montenegro, June 11-14, 2006, Proceedings Series: Lecture Notes in Computer Science, Vol. 4011 Subseries: Information Systems and Applications.

[21] I. Konstas, V. Stathopoulos, and J. M. Jose. On social networks and collaborative recommendation. In SIGIR '09: Proceedings of the 32nd international ACM SIGIR conference on Research and development in information retrieval, pages 195-202, New York, NY, USA, 2009. ACM.

[22] D. Lin. An Information-Theoretic Definition of Similarity. In J. W. Shavlik and J. W. Shavlik, editors, ICML, pages 296-304. Morgan Kaufmann, 1998.

[23] O'Murchu, J.G. Breslin, S. Decker, M. Neumann. Sneachta and the PUII: The semantic social network porta. In Poster at the 3rd International Semantic Web Conference (ISWC 2004), Hiroshima, Japan, Nobember 2004.

[24] M. Newman. Networks An Introduction. Oxford University Press, 2010.

[25] Jordi Palau , Miquel Montaner , Beatriz López , Josep Lluís De La Rosa. Collaboration analysis in recommender systems using social networks. In Cooperative Information Agents VIII: 8th International Workshop, CIA 2004. Volume 3191 of Lectures Notes in Computer Science, pages 137-151, 2004.

[26] Pasquale Lops, Marco Gemmis, and Giovanni Semeraro. Content-based Recommender Systems: State of the Art and Trends. In Francesco Ricci, Lior Rokach, Bracha Shapira, and Paul B. Kantor, editors, Recommender Systems Handbook, chapter 3, pages 73-105. Springer US, Boston, MA, 2011.

[27] Rafter, R., Bradley, K. y Smyth, B. (1999). "Passive Profiling and Collaborative Recommendation." Proc. of the 10th. Irish Conference on Artificial Intelligence and Cognitive Science.

[28] Resnick, P., Iacovou, N., Suchak, M., Bergstrom, P., and Riedl, J., 1994. Grouplens: an open architecture for collaborative filtering of netnews. In CSCW '94: Proceedings of the 1994 ACM conference on Computer supported cooperative work, pages 175-186, New York, NY, USA, 1994.

[29] Sarwar, B., Karypis, G., Konstan, J., and Reidl, J., 2001. Item-based collaborative filtering recommendation algorithms. In Proceedings of the 10th international conference on World Wide Web, WWW '01, pages 285-295, New York, NY, USA, 2001. ACM.

[30] U. Shardanand and P. Maes. Social Information Filtering: Algorithms for Automating \Word of Mouth". In Proceedings of ACM CHI'95 Conference on Human Factors in Computing Systems, volume 1, pages 210-217, 1995.

[31] Jun Wang, Arjen P. de Vries, and Marcel J. T. Reinders. Unifying user-based and item-based collaborative filtering approaches by similarity fusion. In Proceedings of the 29th annual international ACM SIGIR conference on Research and development in information retrieval, SIGIR '06, pages 501-508, New York.

[32] D. Suliema, M. Malek, "Towards Semantic Social recommendation algorithm", in "Seconde conférence sur les Modèles et l′Analyse des Réseaux : Approches Mathématiques et Informatique MARAMI2011." Du 19 au 21 octobre 2011 , Grenoble, France.